\documentclass[twocolumn, superscriptaddress, preview, amsmath, amssymb, aps, letterdraft, notitlepage]{revtex4-1}

\usepackage{graphicx}
\usepackage{amssymb}
\usepackage{color}
\usepackage{booktabs}
\usepackage{subfig}
\usepackage{comment}
\usepackage{appendix}
\usepackage{braket}

\usepackage{color}
\definecolor{KB}{rgb}{0.4,0.3,0.9}

\begin{document}

\title{Fisher information as general metrics of quantum synchronization}

\author{Yuan Shen}
\affiliation{School of Electrical and Electronic Engineering, Nanyang Technological University, Block S2.1, 50 Nanyang Avenue, Singapore 639798}
\author{Hong Yi Soh}	
\affiliation{National Institute of Education,
Nanyang Technological University, 1 Nanyang Walk, Singapore 637616}
\author{Leong-Chuan Kwek}
\email{kwekleongchuan@nus.edu.sg}
\affiliation{School of Electrical and Electronic Engineering, Nanyang Technological University, Block S2.1, 50 Nanyang Avenue, Singapore 639798}
\affiliation{National Institute of Education,
Nanyang Technological University, 1 Nanyang Walk, Singapore 637616}
\affiliation{Centre for Quantum Technologies, National University of Singapore 117543, Singapore}
\affiliation{MajuLab, CNRS-UNS-NUS-NTU International Joint Research Unit, UMI 3654, Singapore}
\author{Weijun Fan}
\email{EWJFan@ntu.edu.sg}
\affiliation{School of Electrical and Electronic Engineering, Nanyang Technological University, Block S2.1, 50 Nanyang Avenue, Singapore 639798}

\def\kwek#1{\textcolor{red}{#1}}

\date{\today}% It is always \today, today,

\begin{abstract}
Quantum synchronization has emerged as a crucial phenomenon in quantum nonlinear dynamics with potential applications in quantum information processing. Multiple measures for quantifying quantum synchronization exist. However, there is currently no widely agreed metric that is universally adopted. In this paper, we propose using classical and quantum Fisher information (FI) as alternative metrics to detect and measure quantum synchronization. We establish the connection between FI and quantum synchronization, demonstrating that both classical and quantum FI can be deployed as more general indicators of quantum phase synchronization, in some regimes where all other existing measures fail to provide reliable results.  We show advantages in FI-based measures, especially in 2-to-1 synchronization. Furthermore, we analyze the impact of noise on the synchronization measures, revealing the robustness and susceptibility of each method in the presence of dissipation and decoherence. Our results open up new avenues for understanding and exploiting quantum synchronization.
\end{abstract}

\maketitle

\section{Introduction}
%introduction

Synchronization is an emergent dynamic process that explains numerous phenomena such as, for instance, the flashing of fireflies ({\it Photinus carolinus}) in tandem \cite{strogatz2012sync}, the clicking of pacemakers \cite{pikovsky2001synchronization}, and the unusual sideward swaying of the Millenium bridge in London \cite{strogatz2005crowd}. A key feature of synchronization is the existence of self-sustaining oscillators coupled to each other or to a driven oscillator.

Synchronization has yielded many interesting  mechanisms for complex systems: limit cycles, amplitude death, oscillation death, and so forth. Synchronization is also intimately connected to chaos theory, where one speaks of the synchronization of chaotic oscillators.
A path less well trampled and studied revolves the synchronization oscillators with co-rotating and counter-rotating orbits in phase space \cite{prasad2010universal, czolczynski2012synchronization, sharma2012counter-rotating,sathiyadevi2022emerging,zeng2011chaos, bhowmick2015counter-rotating,bhowmick2012mixed}.  Such phenomenon has also been known as mixed synchronization in classical literature. 

In recent years, there has been extensive work on quantum versions of classical synchronization \cite{lorch2016phase-coherence,lee2013quantum,walter2014quantum,walter2015quantum}. Instead of the classical phase space, one investigates the Wigner function and probes into the presence of an Arnold-like tongue.  For more than one oscillator, measures have been devised to detect the presence of quantum synchronization \cite{jaseem2020generalized,mari2013measures}. As in the classical case, we know less about quantum mixed synchronization. In the classification of measures for continuous variable quantum system, the authors have briefly mentioned mixed synchronization for two oscillators with opposite momenta and its deep connection to Einstein-Podolsky-Rosen pairs\cite{mari2013measures}.  

Moreover, in the quantum regime, it is known that a limit-cycle oscillator with a squeezing Hamiltonian can undergo a bifurcation where the Wigner function splits into two symmetrical peaks~\cite{sonar2018squeezing}. In Ref. \cite{weiss2016noise}, the authors have investigated two quantum oscillators that display peak synchronization at two different phases, with a phase of $\pi$ apart. In such cases, the system can be regarded as in-phase synchronization with the drive, but the phase locking happens at two distinct phases.  We refer to this phenomenon as 2-to-1 synchronization, and it is similar to the case described as mixed synchronization in Weiss {\it et al.} ~\cite{weiss2016noise}.

Fisher information has been used as a measure of the ability to estimate an unknown parameter or as a measure of the state of disorder of a system \cite{frieden1990fisher,frieden2004science}. The quantum Fisher information (QFI) serves as a crucial measure in quantum parameter estimation, providing insights into the precision with which a quantum system can estimate an unknown parameter. 

In this paper, we look at various measures for quantum synchronization of an externally driven oscillator and explore the possibility of introducing Fisher information to define quantum synchronization, especially for the general case of $n$-to-1 quantum synchronization. This paper is organized as follows: in Section \ref{sec:model}, we describe the driven quantum Stuart-Landau oscillator and we discuss various possible measures of synchronization. In Section \ref{sec:phase_synch}, we study the behavior of the various measures of synchronization and discuss 2-to-1 synchronization in Section \ref{squeezed}, the case where a squeezing term is added to the oscillator.  There are different possible noises in such system, we investigate the effects of noise in Section \ref{sec:noise}. In section \ref{sec:different_measures}, we compare the correlations between the different measures. Finally, in section \ref{sec:asymmetric}, we investigate the asymmetric case of 2-to-1 synchronization and make some concluding remarks in section \ref{sec:conclusion}.

\section{Oscillator model and synchronization measures}\label{sec:model}

We study the quantum van~der~Pol oscillator (also known as the quantum Stuart-Landau oscillator~\cite{shen2023quantum_sync_nonlinear}) subjected to both single photon drive and two photon squeezing drive. The master equation in the rotating frame of the drive gives~(with $\hbar = 1$):
\begin{align}\label{eqn:master_eq}
    \dot \rho = &-i[\hat{H},\rho] + \gamma_1 \mathcal{D}[a^\dagger]\rho +\gamma_2\mathcal{D}[a^2]\rho 
    +\gamma_3\mathcal{D}[a]\rho\nonumber\\
    \hat{H} = & \Delta a^\dagger a + iE(a-a^\dagger) + i \eta(a^{\dagger 2} e^{2i\varphi} - a^{2} e^{-2i\varphi}) ,
\end{align}
where $\mathcal{D}[L]\rho = L\rho L^\dagger - \frac{1}{2}(L^\dagger L \rho + \rho L^\dagger L)$, $\gamma$ represents the rate of decay, with $\gamma_1$, $\gamma_2$ and $\gamma_3$ corresponding to negative damping, nonlinear damping and linear damping respectively.  $\Delta = \omega_0-\omega_d$ is the amount of detuning between the frequency of the drive, $\omega_d$, and the frequency of the oscillator, $\omega_0$. $E$ is the amplitude of the harmonic drive, with $a$ and $a^\dagger$ being the annihilation and creation operators. $\eta$\ is the squeezing parameter, with $\varphi$ representing the phase of squeezing. 

In this paper, we focus on the measures of quantum phase synchronization in an externally driven oscillator.  As a measure of phase synchronization, the phase coherence is frequently used in the literature and, defined as~\cite{weiss2016noise,lorch2016phase-coherence,barak2005non,shen2023enhance_homodyne}
\begin{equation}
    S_{pcoh}  = \frac{\mbox{Tr}[a\rho]}{\sqrt{\mbox{Tr}[a^\dagger a\rho]}},
\end{equation}
where $|S|$ measures the degree of phase coherence with a range of $0\le |S|\le 1$. Another appropriate simple measure is based on the relative phase distribution~\cite{hush2015spin_correlations,lorch2017sync_blockade}:
\begin{equation}
    S_{peak} = 2\pi \  \mbox{\rm max}[P(\Phi)]-1,
\end{equation}
where the phase distribution is defined by $P(\Phi)=(1/2\pi)\langle\Phi|\rho|\Phi\rangle$ with $|\Phi\rangle=\sum_{n=0}^{\infty}e^{in\Phi}|n\rangle$. $S_{peak} $ represents the maximum value of $P(\Phi)$ compared to a uniform distribution. This measure is valuable for detecting synchronization because it is exclusively nonzero when $P(\Phi)$ deviates from a flat distribution.

It is well known that the phase operator is not well-defined in quantum theory. However, most quantum harmonic oscillators are populated up to some finite levels, and we can resort to the Pegg-Barnett phase operator. In Ref.~\cite{mok2020sync_boost}, the mean resultant length~($MRL$), which incorporates the Pegg-Barnett operator, has been proposed as a measure of synchronization. It arises from the study of circular statistics \cite{pewsey2013circular} and is initially developed for 1-to-1 synchronization. However, it can be generalized to measure n-to-1 synchronization. The \textit{n}-th order mean resultant length~($MRL^{(n)}$) of a circular distribution is given by
\begin{equation}
    MRL^{(n)}=\sqrt{\langle \sin n\phi \rangle^2 + \langle \cos n \phi \rangle^2} = |\langle e^{i n\phi} \rangle|.
\end{equation}
This measure is capable of capturing $n$-to-1 synchronization, which exhibits multiple peaks in the phase distribution $P(\Phi)$ and fixed-points in the quasi-probability phase-space distribution, e.g. Wigner function. 

Fisher information proves to be an important tool for determining classical synchronization in a system of Kuramoto oscillators \cite{kalloniatis2018fisher, da2021fisher}.  It has been mooted as a good measure for phase drift in clock synchronization, both classical and quantum \cite{yue2015operation,zhang2013criterion,jozsa2000quantum,chen2010clocks}.  It is also a useful measure in classical signal processing \cite{steven1993fundamentals}, being intimately related to the Cramer-Rao bound. 
Motivated by these works, we propose the quantum Fisher information~(QFI) as a measure for quantum phase synchronization: 
\begin{equation}\label{eqn:qfi}
    QFI=\mathcal{F}_Q[\rho,A] = 2 \sum_{k,l} \frac{(\lambda_k-\lambda_l)^2}{(\lambda_k+\lambda_l)} |\langle k|A|l\rangle |^2 ,
\end{equation}
where $\lambda_{k,l}$ and $|k,l\rangle$ are the eigenvalues and eigenvectors of the steady-state $\rho = \rho_{ss}$. We use $A = a^\dagger a$ to measure the phase uncertainty in the steady-state. 

Phase synchronization is closely related to the phase distribution $P{(\Phi)}$, which is a classical probability distribution. Therefore, it make sense to directly inspect this classical distribution to obtain information about synchronization. We propose another measure of phase synchronization using classical Fisher information~(CFI). This new measure is defined by the classical Fisher information of the phase distribution $P(\Phi)$:
\begin{align}
    CFI = E\left[\big(\frac{\partial}{\partial \Phi}\log P(\Phi)\big)^2\right],
\end{align}
It is important to note that this CFI is different from the conventional Fisher information, which is directly calculated from the density matrix as: $F(\hat X|\theta)=\sum_x \frac{1}{p(x|\theta)}(\frac{\partial p(x|\theta)}{\partial \theta})^2$, where $p(x|\theta)$ is the probability of observing outcome $x$ when measuring observable $\hat X$~\cite{paris2009quan_estimate}.

Classical and quantum Fisher information and $S_{peak}$ reads $0$ for unsynchronized states, but is unbounded for highly synchronized states, whereas phase coherence and $MRL^{(n)}$ is bounded between $0$ and $1$.

\begin{figure}[t]
    \centering
    \includegraphics[width=\linewidth]{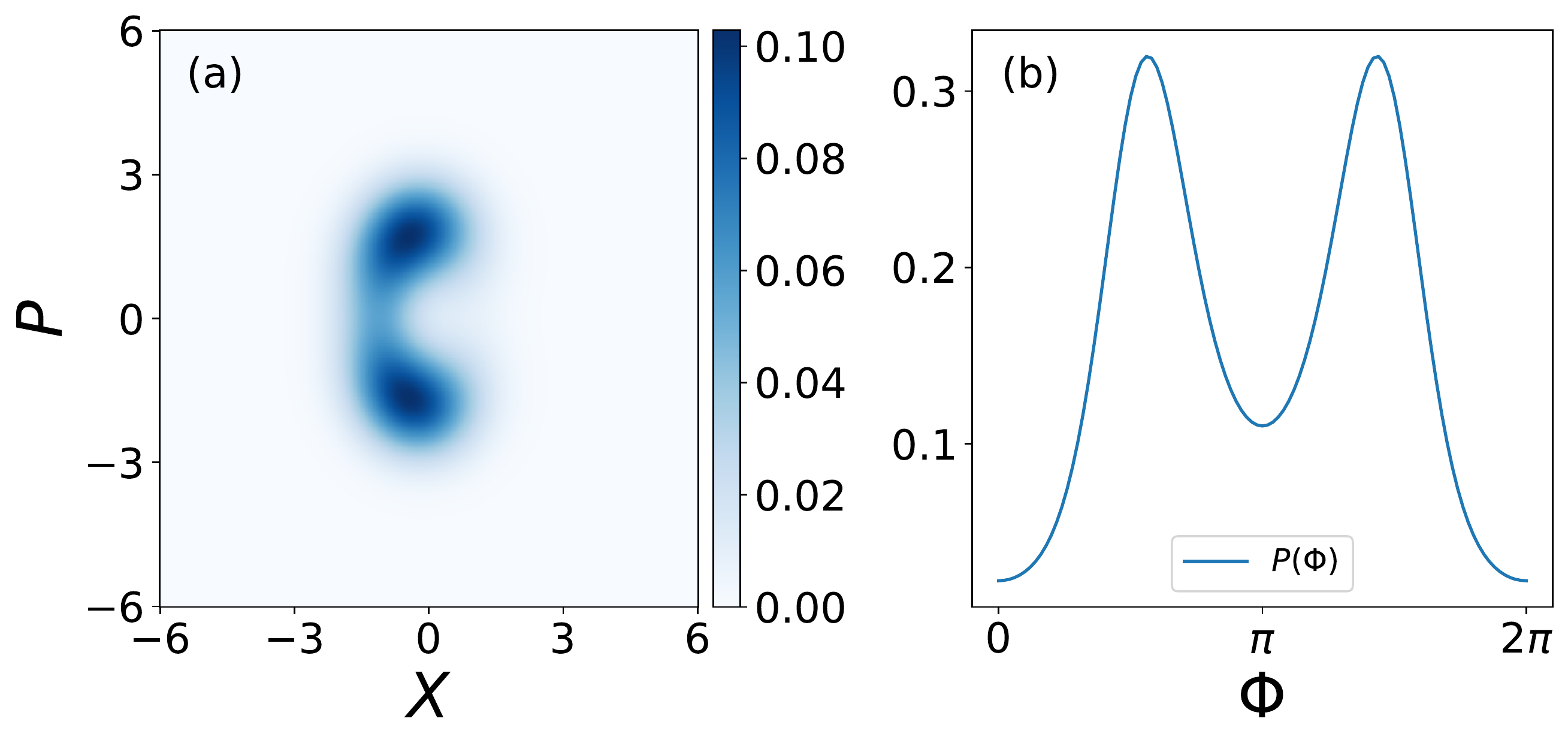}
    \caption{An example of a squeezed steady-state (a) Wigner function and its corresponding (b) phase distribution $P(\Phi)$. Squeezed Wigner function and phase distribution have two distinct peaks, which we refer to as 2-to-1 synchronization. Parameters in this example: $\Delta=0,E=\eta=0.5,\varphi=\pi/2,\gamma_1=\gamma_2=1,\gamma_3=0$.}
   \label{fig:fig1}
\end{figure}

\begin{figure}[t]
    \centering
    \includegraphics[width=\linewidth]{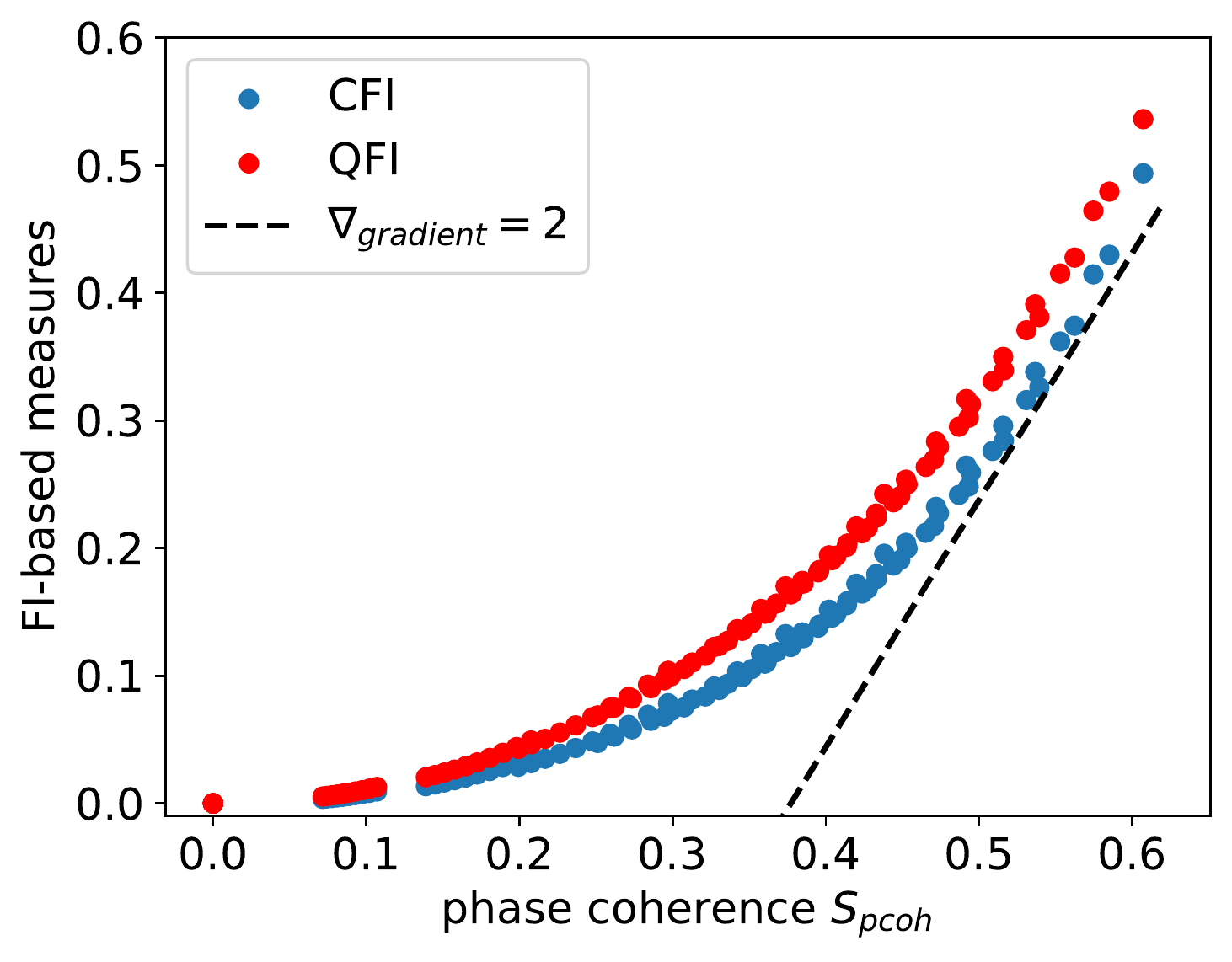}
    \caption{Phase coherence vs. FI-based measures. It is interesting to see FI-based measures is more sensitive(larger gradient) for highly synchronized states, as shown by the dotted reference line indicating $\nabla_{gradient}=2$. Sample data simulated with: $\Delta=0,E=0.5,\eta=0,\varphi=\pi/2,\gamma_1=1,\gamma_2\in [1,10],\gamma_3=0$.}
    \label{fig:fig2}
\end{figure}

Two advantages of FI-based measures over phase coherence can be observed: Firstly, Fisher information appears to be more sensitive to highly synchronized states, while exhibiting less sensitivity at the other extreme.  However, in most cases, our primary interest lies in the highly synchronized states. Secondly, FI-based measures are more general metrics of synchronization. Measures such as phase coherence face limitations in detecting synchronization of squeezed states or, more generally, Wigner functions with multiple peaks— $n$-to-1 synchronization, see Fig.~\ref{fig:fig1} as an example. In contrast, FI-based measures are capable of detecting synchronization in such instances. As a measure of synchronization, we find that FI-based measures are not only comparable to existing measures for normal cases of 1-to-1 synchronization, it is also more appropriate for the measurement of 2-to-1 synchronization. Measuring QFI in experiments can be challenging due to its reliance on the full quantum state of the system. However, there are various strategies that have been developed to estimate QFI experimentally, such as randomized measurement~\cite{rath2021qfi_rand_meas,frowis2016detect_qfi,yu2021exp_est_qfi}.

Recently there has been some work \cite{daniel2023geometric} to relate quantum synchronization to quantum geometric phase \cite{tong2004kinematic}.  In this work, they showed that the geometric phase for the quantum Stuart-Landau  oscillator under driven pump exhibits an Arnold tongue-like structure, somewhat similar to the Arnold tongue in quantum synchronization as measured by the shifted phase distribution of the $Q$ function.  Also, for two oscillators, it is sometimes useful to measure the  quantum mutual information \cite{ameri2015mutual,jaseem2020generalized, eneriz2019degree}.

\section{1-to-1 Synchronization}\label{sec:phase_synch}

\begin{figure}[t]
    \centering
    
    \includegraphics[width=\linewidth]{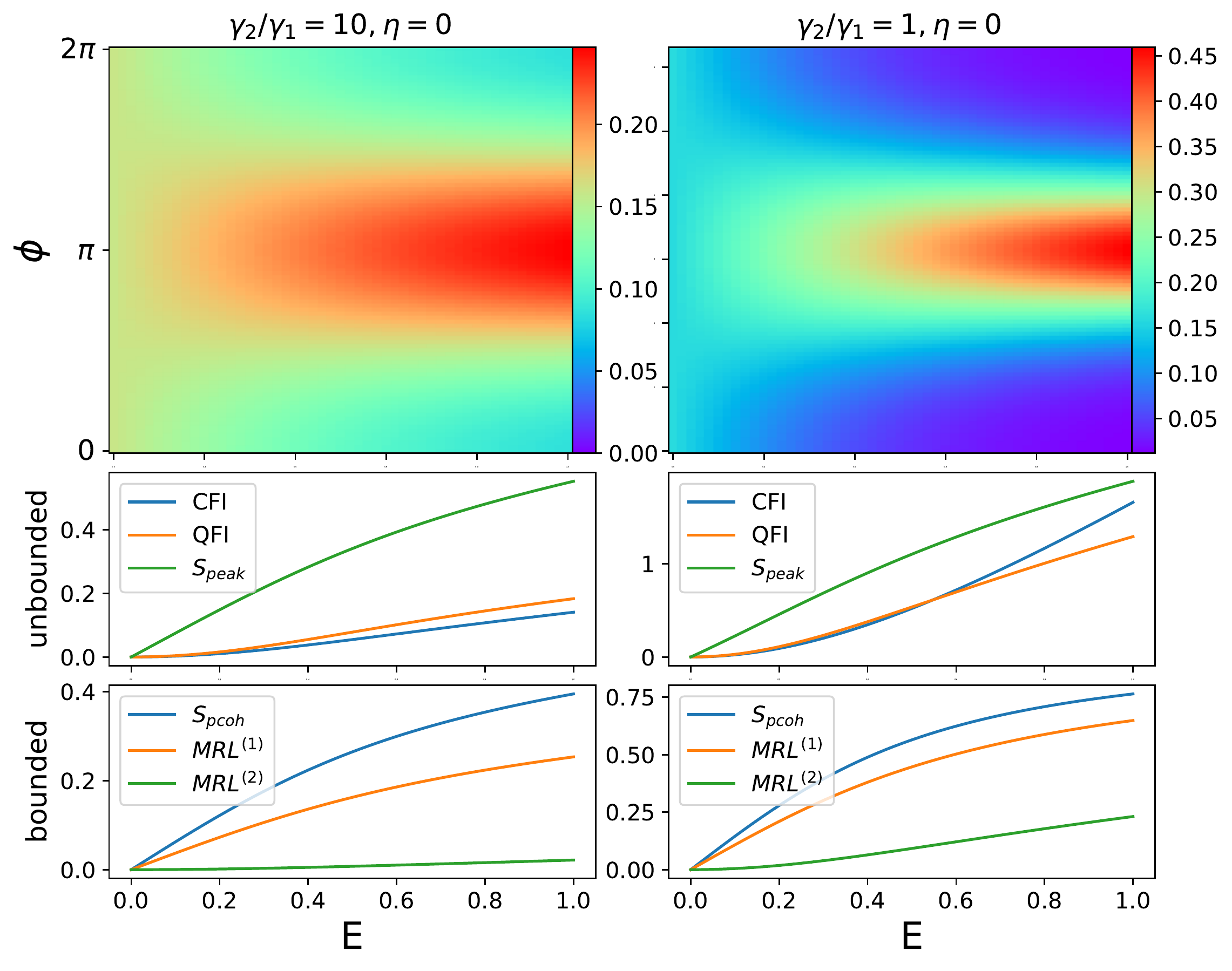}
    \caption{Phase distribution $P(\phi)$ and synchronization measures plotted against driving amplitude $E$. Fixed parameters: $\Delta=0,\gamma_1=1,\gamma_3=0$. In these cases of $1$-to-$1$ synchronization, the driven oscillator has only one preferred phase to synchronize to. Unbounded and bounded measures are plotted separately.}
    \label{fig:fig3}
\end{figure}

We first study the scenario of a coherently driven oscillator without squeezing drive(by simply set $\eta=0$). When only the coherent driving is present, there will be only one preferred phase~(namely 'fixed point') to synchronize to  and the phase distribution $P(\Phi)$ has only one peak, as shown by the first row in Fig.~\ref{fig:fig3}, whose position indicates the relative phase between the oscillator and drive. With increasing amplitude of the driving, the quantum phase synchronization between the oscillator and drive improves, and so does the values of the synchronization measures. This indicates a monotonic behavior in the measure. We show that all measures qualitatively agree in Fig.~\ref{fig:fig3} where synchronization measures are plotted against coherent driving amplitude $E$. Therefore, these are all valid measures to capture 1-to-1 quantum phase synchronization, and their correlations are close to unity, as shown in the later section. Take note that in Fig.~\ref{fig:fig3} the unbounded and bounded measures are plotted separately. 

In Fig.~\ref{fig:fig3}, the synchronization measures are compared across different nonlinear damping ratios $\gamma_2/\gamma_1$, where this ratio directly controls the radius of the limit cycle and mean photon number in the oscillator. Conventionally, the oscillator is regarded in 'semi-classical' regime when $\gamma_2/\gamma_1\approx 1$, and 'quantum' regime when $\gamma_2/\gamma_1\gg 1$. We can see that these measures remain valid for different regimes. A driven oscillator with a smaller radius~(i.e. larger $\gamma_2/\gamma_1$) is more prone to lose synchronization by phase diffusion and quantum noise~\cite{lorch2016phase-coherence,lee2013quantum,walter2014quantum}. Comparing two columns of Fig.~\ref{fig:fig3}, the values of a synchronization measure are higher in the classical regime, as expected. Note that in Fig.~\ref{fig:fig3} right column, the value of CFI surpass QFI at certain driving amplitude $E$. As we have explained previously, the CFI we proposed in this paper is not the direct classical counter-part of QFI. Therefore, this is not a violation of the property  that QFI should be the supremum of the CFI over all observables.

More insights can be developed in deep quantum regime~($\gamma_2 \rightarrow \infty$), where the analytical solutions to all these measures can be obtained. By using the $3\times 3$ density matrix ansatz proposed in~\cite{mok2020sync_boost}, the analytical equation for $MRL^{(1)}$, QFI and phase coherence $S_{pcoh}$ are obtained as  (with $\Delta=0,\gamma_1=1,\gamma_3=0$):
\begin{equation}
    \lim_{\gamma_2 \to \infty}MRL^{(1)}=\frac{2E}{9+8E^2},
\end{equation}

\begin{equation}
    \lim_{\gamma_2 \to \infty}QFI = 4 \bigg|\frac{2E}{9+8E^2}\bigg|^2,
\end{equation}

\begin{equation}
    \lim_{\gamma_2 \to \infty}S_{pcoh} = \frac{2E}{\sqrt{(8E^2+9)(4E^2+3)}}.
\end{equation}

Subsequently, the phase distribution $P(\Phi)$ can be obtained:
\begin{equation}
    \lim_{\gamma_2 \to \infty} P(\Phi)=\frac{1}{2\pi}[1-\frac{4E}{9+8E^2}\cos(\Phi)]
\end{equation}

After deriving the phase distribution $P(\Phi)$, the peak of phase distribution $S_{peak}$ and CFI can be easily obtained as 
\begin{equation}
    \lim_{\gamma_2 \to \infty}S_{peak} = \frac{4E}{9+8E^2} ,
\end{equation}

\begin{equation}
    \lim_{\gamma_2 \to \infty}CFI = 4\frac{A_0 + A_1 E^2 +A_2 E^4 + A_3 E^6 +A_4 E^8}{\lambda (9+8E^2)(\lambda-9-8E^2)^2} ,
\end{equation}
where 
\begin{align}
    A_0 &= 729(\lambda-9),\nonumber\\ A_1 &= 108(17\lambda -201), \nonumber\\
    A_2 &= 544(3\lambda - 52), \nonumber\\A_3 &= 256(2\lambda - 67), \nonumber\\
    A_4 &= -4096,\nonumber\\
    \lambda &= \sqrt{(9+4E+8E^2)(9-4E+8E^2)}.
\end{align}

These solutions are valid when driving amplitude $E<<1$~(see Appendix A), beyond which the density matrix ansatz breaks down.

\section{Squeezing enhances 2-to-1 synchronization}\label{squeezed}

\begin{figure}[t!]
    \centering
    \includegraphics[width=\linewidth]{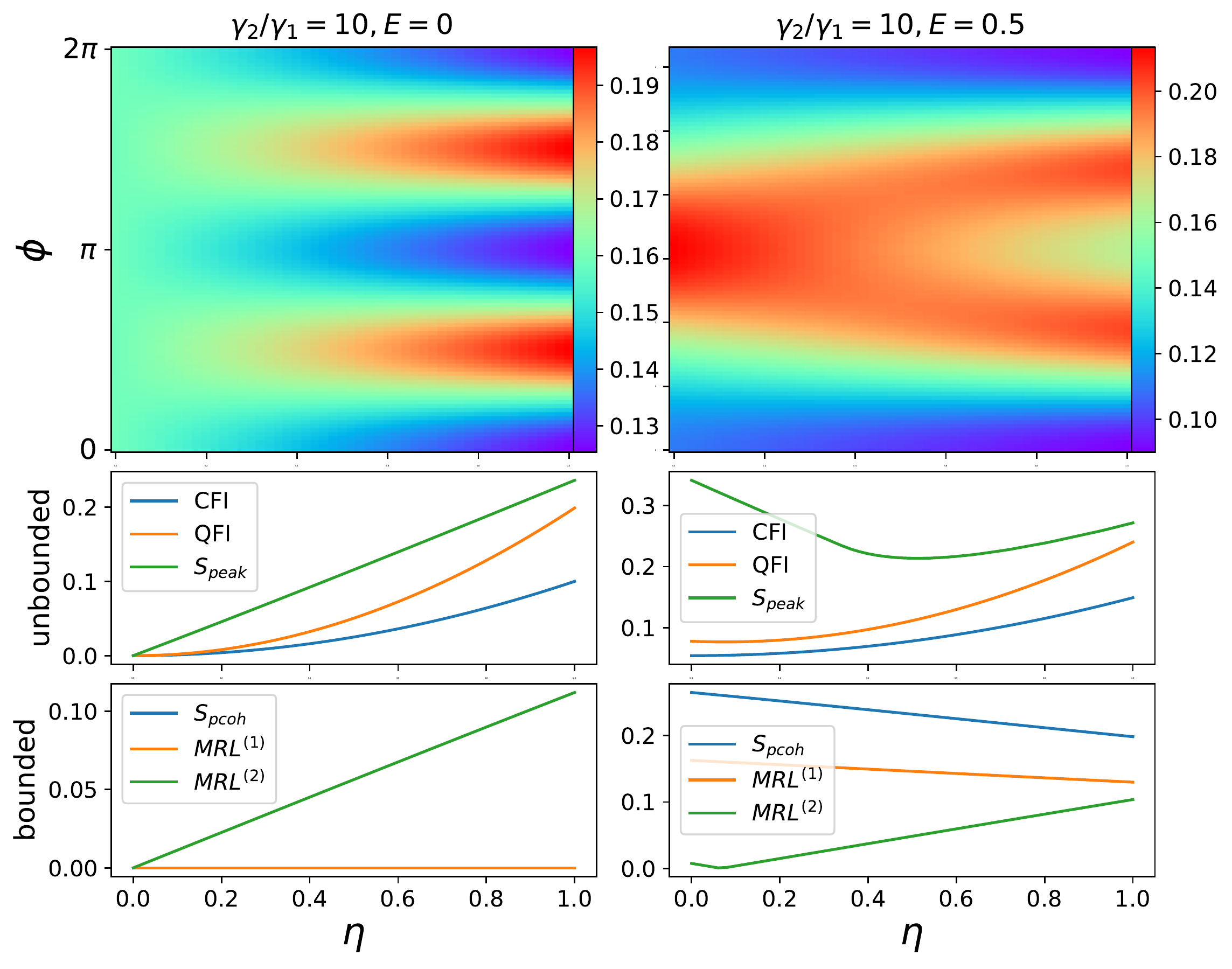}
    \caption{Phase distribution $P(\phi)$ and synchronization measures plotted against squeezing amplitude $\eta$. Fixed parameters: $\Delta=0,\varphi=\pi/2,\gamma_1=1,\gamma_3=0$. In these cases of $2$-to-$1$ synchronization, the driven oscillator has two distinct phase to synchronize to. Unbounded and bounded measures are plotted separately.}
    \label{fig:fig4}
\end{figure}

 In this section, we show squeezing drive can create and enhance quantum 2-to-1 synchronization, i.e. synchronization with two distinct fixed points in phase space, and the phase distribution $P(\Phi)$ has two distinct peaks. However, as mentioned previously, some synchronization measures are not suitable for measuring such type of synchronization.

As shown in Fig.~\ref{fig:fig4} left column, increasing squeezing sharpens the two peaks in the phase distribution and thus improves mixed synchronization. Here, we need to consider the following question: Does $E=0$. i.e. no drive, makes sense for synchronization?  We can always regard the squeezing term as a drive. When squeezing is present without coherent drive, the synchronization can be regarded as between the oscillator and the squeezing drive. This is also considered in Ref.~\cite{sonar2018squeezing} where they did for frequency entrainment~(the frequencies of the oscillator and external drive converge). Note that phase coherence $S_{pcoh}$ and $MRL^{(1)}$ is zero when only squeezing is present, which is expected, as these measures reflect the first off-diagonal elements in the density matrix. On the other hand, $S_{peak}$ and $MRL^{(2)}$ scale almost linearly with squeezing.

2-to-1 synchronization can be created out of 1-to-1 synchronization. This is shown in the right column of Fig.~\ref{fig:fig4}, where in addition to squeezing, a coherent drive with amplitude $E=0.5$ is present. This coherent drive creates a single peak when squeezing is off or small. When the squeezing is tuned up, the single peak splits into two under pitchfork bifurcation, so does the corresponding Wigner functions~\cite{sonar2018squeezing}. In this scenario, the two measures (phase coherence $S_{pcoh}$ and $MRL^{(1)}$) which are only capable of measuring 1-to-1 synchronization decrease and appears to change almost linearly with increasing squeezing parameter.   
$MRL^{(2)}$ is a measure dedicated to 2-to-1 synchronization, therefore it is unsurprising that it only provides partial information when single peak is present. This explains why $MRL^{(2)}$ drops to zero at small $\eta$ and increases linearly afterwards. The measurement of $S_{peak}$ lacks the ability to differentiate between two types of synchronization. Consequently, only the classical and quantum Fisher information measures exhibit a monotonic relationship with respect to squeezing.

\section{Effect of noise}\label{sec:noise}
In this section, we investigate and compare the effect of different noise across these measures. We consider two types of noise, namely single photon dissipation and white noise.

\begin{figure}[t!]
    \centering
    \includegraphics[width=\linewidth]{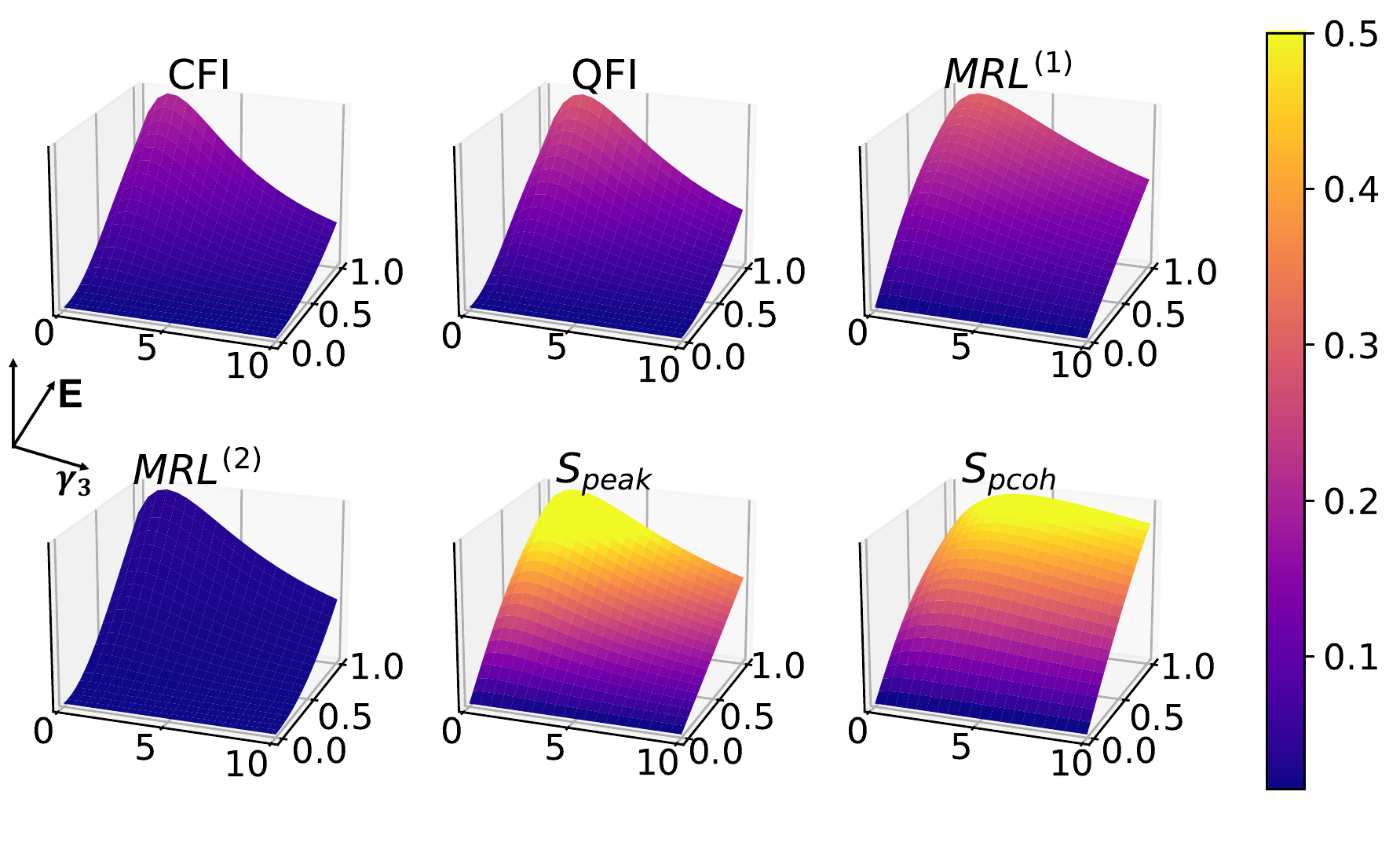}
    \caption{Effects of single-photon dissipation noise in 1-to-1 synchronization, in the absence of squeezing~($\eta=0$), with fixed parameters: $\Delta=0,\gamma_1=1,\gamma_2=10,p=0$. All measures exhibit an noise-induced boost where the dissipation is small, which is consistent with the previous works~\cite{mok2020sync_boost,shen2023enhance_homodyne}.}
    \label{fig:single_photon_noise1}
\end{figure}

\begin{figure}[hbt]
    \centering
    \includegraphics[width=\linewidth]{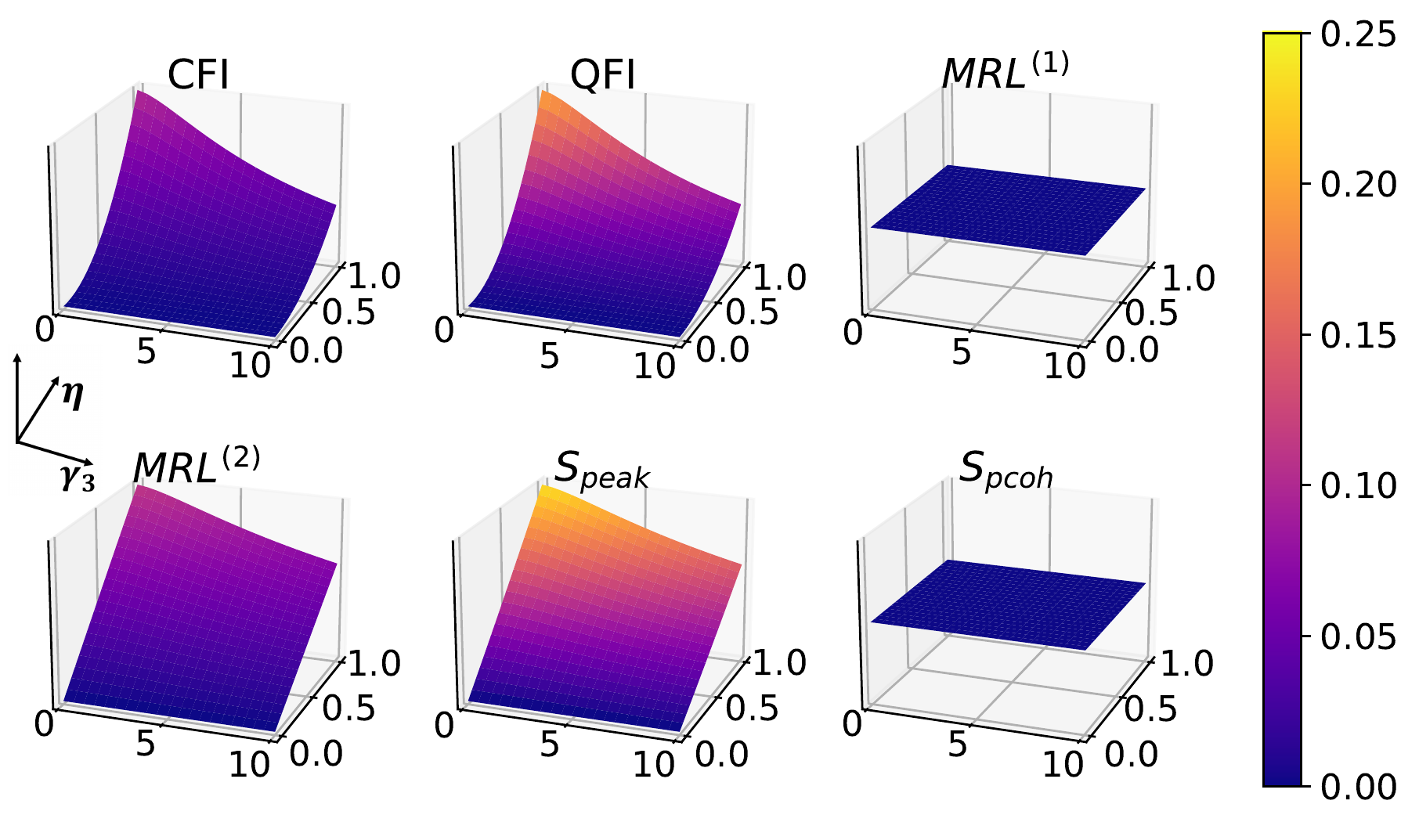}
    \caption{Effects of single-photon dissipation noise in 2-to-1 synchronization, without driving~($E=0$), with fixed parameters: $\Delta=0,\gamma_1=1,\gamma_2=10,p=0$. As discussed above, $MRL^{(1)}$ and phase coherence are identically zero in these cases.}
    \label{fig:single_photon_noise2}
\end{figure}

The single photon dissipation process is implemented by the Lindblad dissipator proportional to $\gamma_3$ in the master equation~\eqref{eqn:master_eq}. In Fig.~\ref{fig:single_photon_noise1}, all six measures are captured in the surface plots with respect to the single photon dissipation $\gamma_3$ and coherent driving amplitude $E$. It is known that single photon dissipation can be beneficial for 1-to-1 synchronization in coherently driven oscillators~\cite{mok2020sync_boost,shen2023enhance_homodyne}, which is reflected in Fig.~\ref{fig:single_photon_noise1} among all measures consistently. Surprisingly, this noise-induced synchronization boost is absent in 2-to-1 synchronization, as shown in Fig.~\ref{fig:single_photon_noise2}, in which the squeezing $\eta$ is increasing instead of driving amplitude $E$. Again, the phase coherence and $MRL^{(1)}$ remain $0$ for the same reason explained in the previous section.

To introduce white noise into the density matrix, we define noise parameter $p\in [0,1]$, thus the noisy steady-state density matrix is defined as:
\begin{equation}
    \rho_{noisy} = (1-p) \rho_{ss}  + p I/N_{dim} ,
\end{equation}
where $\rho_{ss}$ is the noise-less steady-state density matrix and $I$ is the identity matrix with dimension $N_{dim}$.

\begin{figure}[t!]
    \centering
    \includegraphics[width=\linewidth]{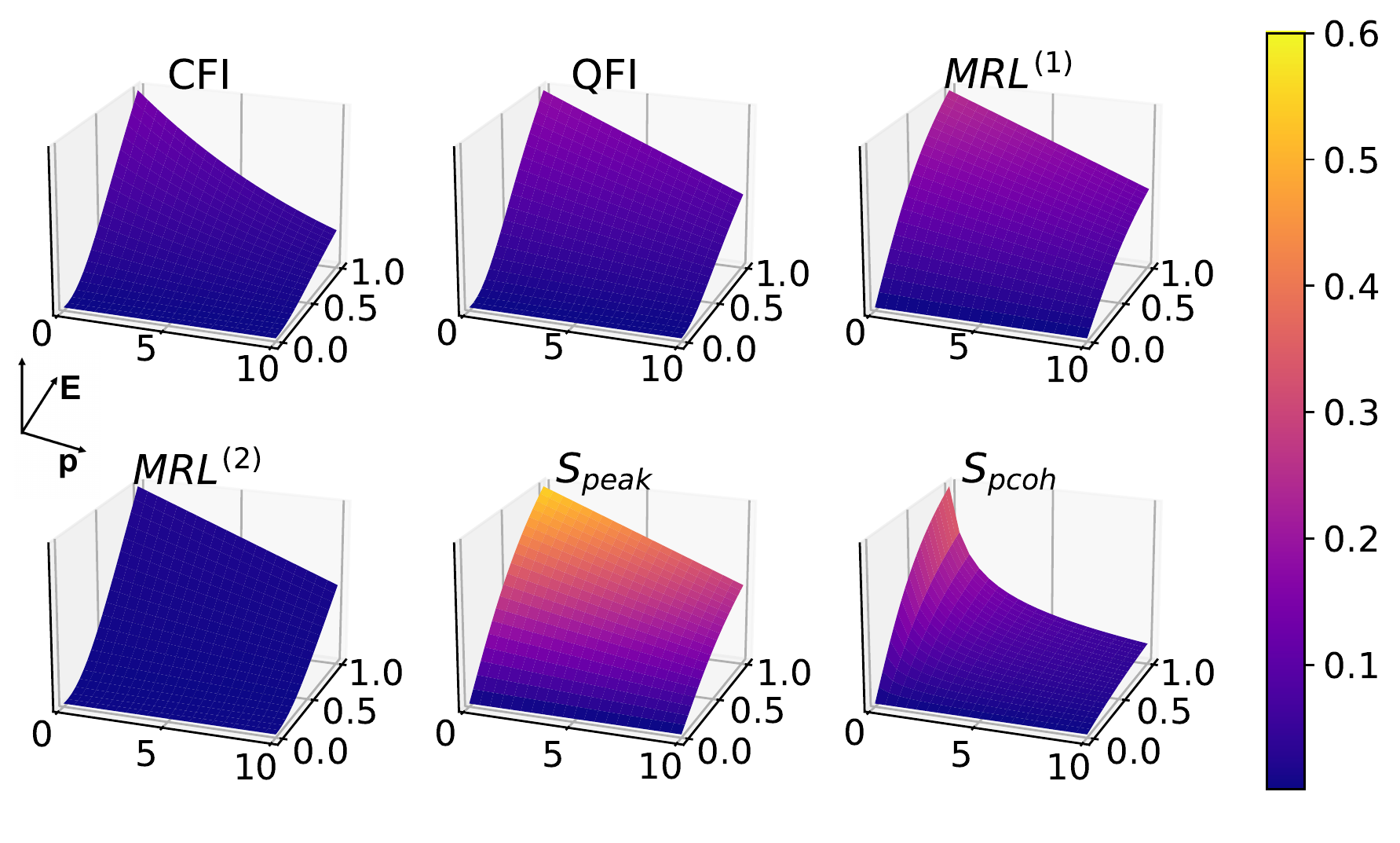}
    \caption{Effects of white noise in 1-to-1 synchronization, in the absence of squeezing~($\eta=0$), with fixed parameters: $\Delta=0,\gamma_1=1,\gamma_2=10,\gamma_3=0$.}
    \label{fig:white_noise1}
\end{figure}

After introducing white noise, it is expected for all measures to degrade with increasing $p$. Interestingly, phase coherence turns out to be most sensitive to white noise, as shown in Fig.~\ref{fig:white_noise1}, where there is a bigger drop in the measure as a function of noise $p$ compared to other measures.

\begin{figure}[hbt]
    \centering
    \includegraphics[width=\linewidth]{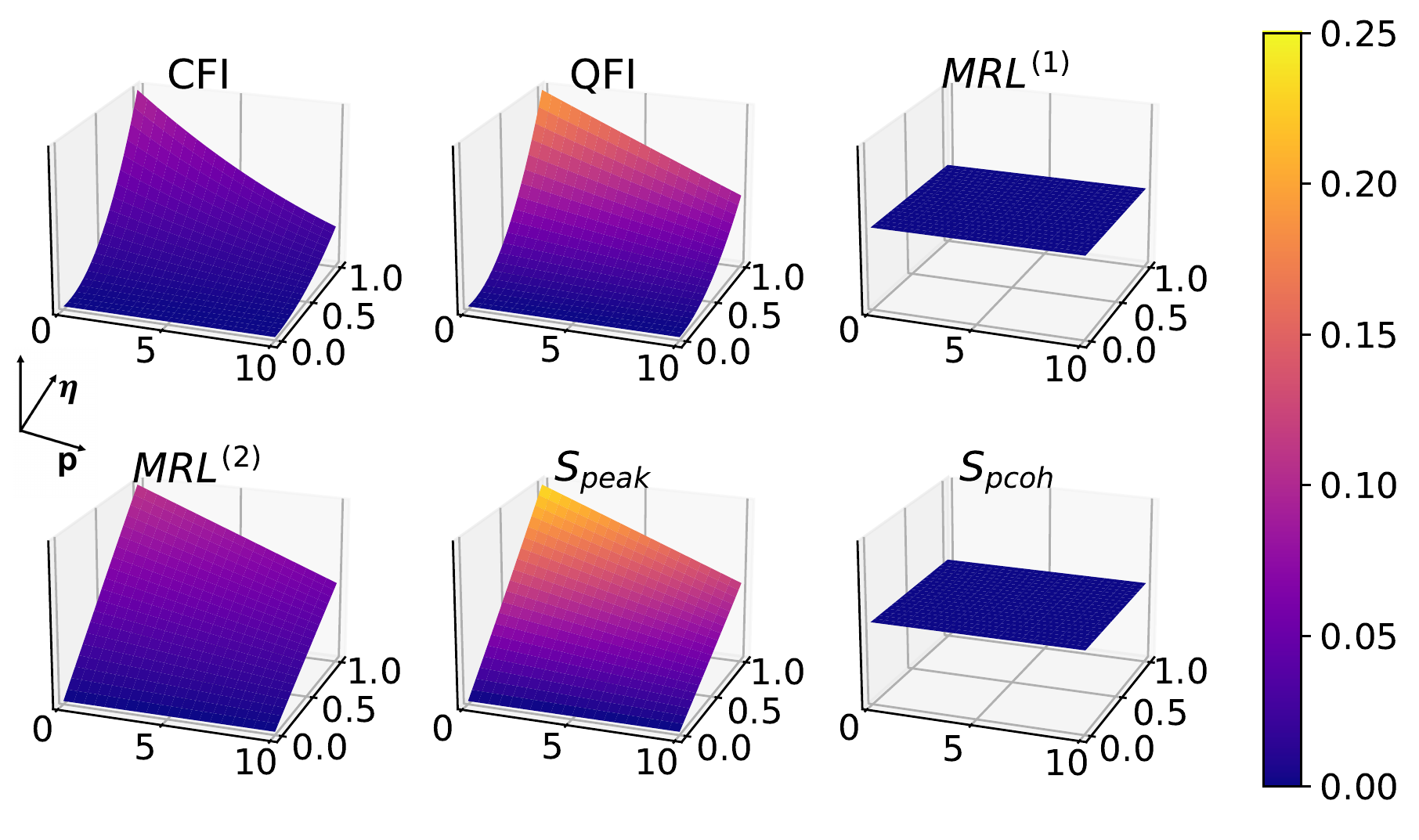}
    \caption{Effects of white noise in 2-to-1 synchronization, without driving~($E=0$), with fixed parameters: $\Delta=0,\gamma_1=1,\gamma_2=10,\gamma_3=0$. $MRL^{(1)}$ and phase coherence are $0$ in this case for the same reason as above.}
    \label{fig:white_noise2}
\end{figure}

\section{Correlations between measures}\label{sec:different_measures}

In this section, a correlation analysis was performed to investigate to what extent the different measures of quantum synchronization carry independent and non-redundant information. We calculate the Pearson correlation between the values of different measures, defined as 
\begin{equation}
    \mathcal{C} = \frac{\mbox{cov}(X,Y)}{\sigma_X \sigma_Y},
\end{equation}
with $\mbox{cov}(X,Y)$ being the covariance between two synchronization measures and $\sigma$ the standard deviation.
 In the case of 1-to-1 synchronization, i.e. single peak, high Pearson correlations are observed across all the measures, as shown in Fig.9. Whereas in the case of 2-to-1 synchronization, it is obvious that phase coherence $S_{pcoh}$, $S_{peak}$ and $MRL^{(1)}$ are ill-suited measures, as they are negatively related to the other three proper measures.

 From both plots we can tell the connections between these measures: CFI, QFI and $MRL^{(2)}$ are highly correlated in their response to the driving. On the other hand, phase coherence $S_{pcoh}$ and $MRL^{(1)}$ exhibit a strong connection, as they are both related to the first off-diagonal coherences. 

\begin{figure}[ht]
    \centering
    \includegraphics[width=0.8\linewidth]{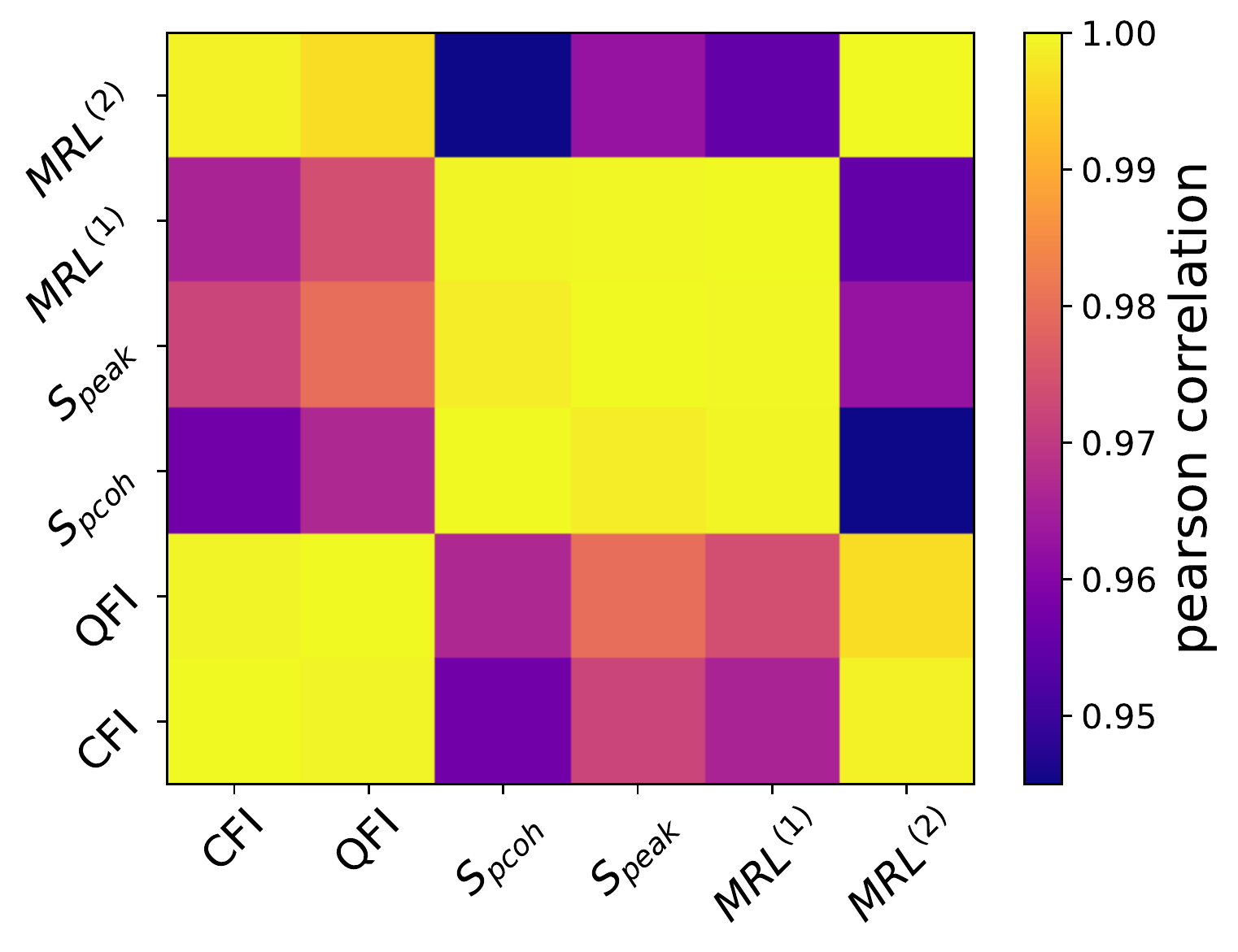}
    \caption{Correlation between different measures on 1-to-1 synchronization. Calculations are performed on the same data as Fig.~\ref{fig:fig3} left column.}
    \label{fig:pearson_corr_1}
\end{figure}

\begin{figure}[ht]
    \centering
    \includegraphics[width=0.8\linewidth]{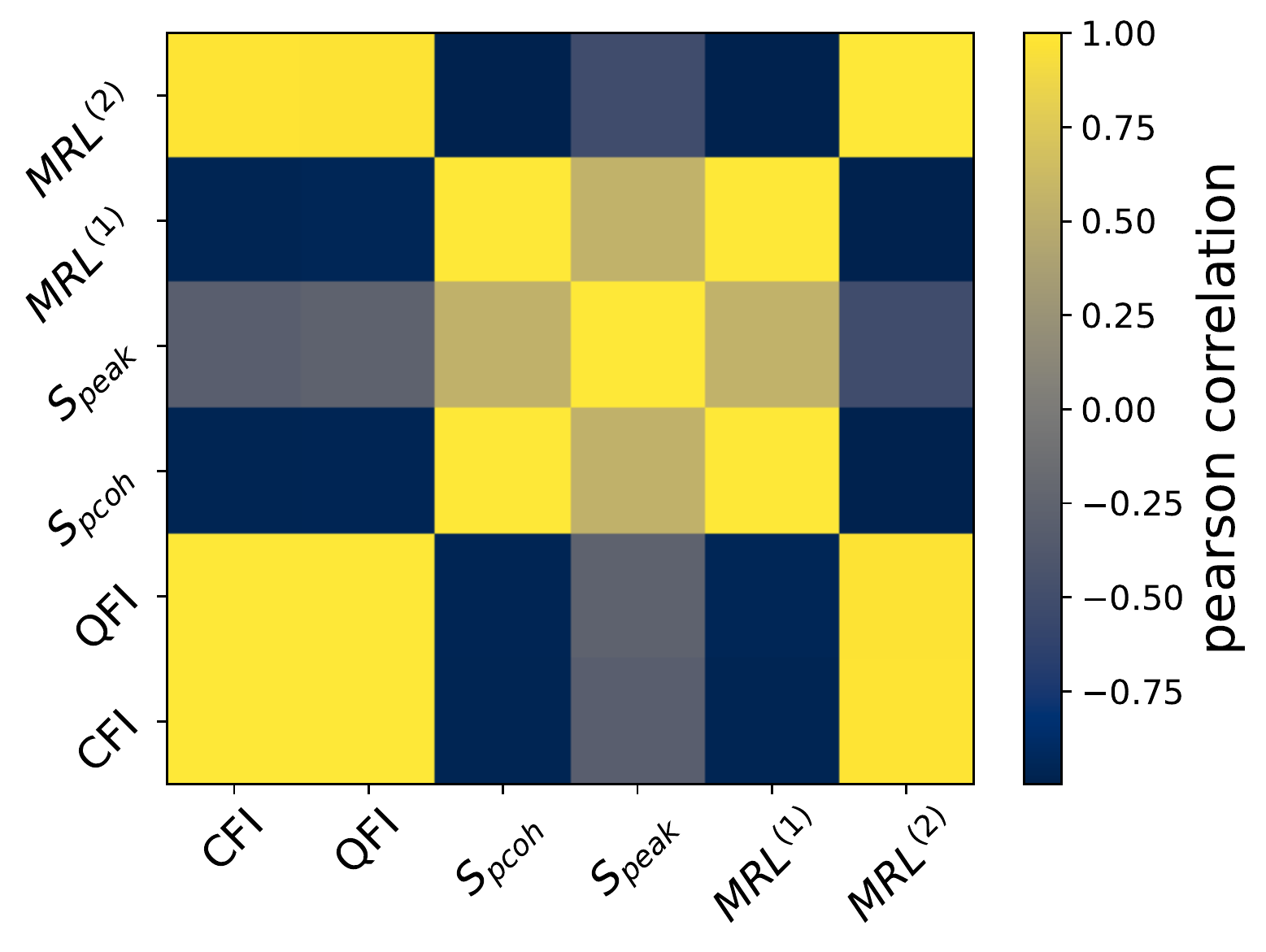}
    \caption{Correlation between different measures on 2-to-1 synchronization. Calculations are performed on the same data as Fig.~\ref{fig:fig4} right column.}
    \label{fig:pearson_corr_2}
\end{figure}

\section{Asymmetrical synchronization}\label{sec:asymmetric}

\begin{figure}[ht!]
    \centering
    \includegraphics[width=\linewidth]{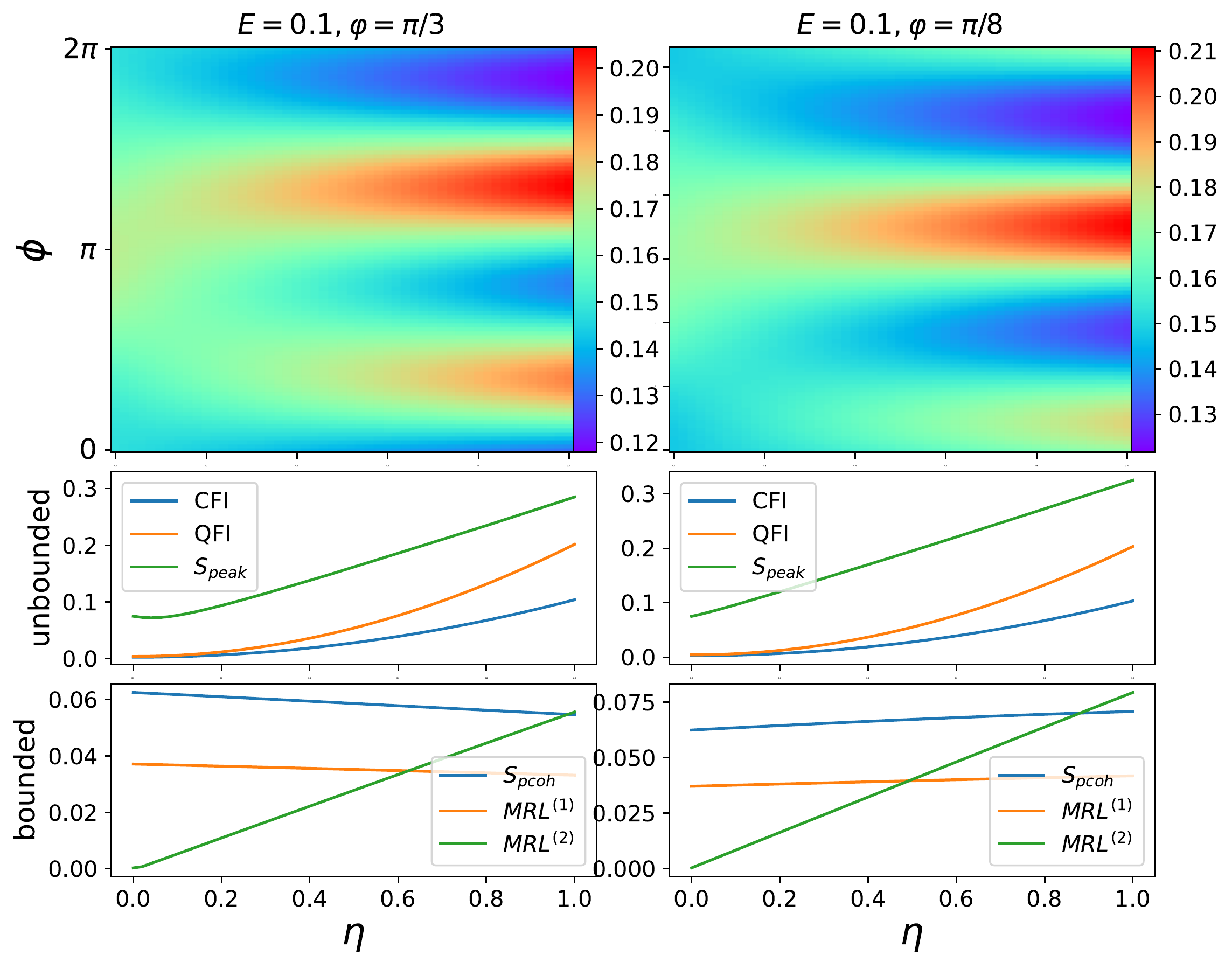}
    \caption{Asymmetrical phase distribution and synchronization measures plotted against squeezing $\eta$. Fixed parameters: $\Delta=0,\gamma_1=1,\gamma_2=10,\gamma_3=0$.}
    \label{fig:asymmetric}
\end{figure}

So far we have discussed cases when the two peaks in phase distribution are symmetrical~(i.e. the peaks have identical amplitude). To complete the whole picture, in this section we discuss the situation when the two peaks are distorted and asymmetrical. This asymmetrical phase distribution can be observed when both coherent drive and squeezing are present with a difference of phase, as shown in Fig.~\ref{fig:asymmetric}. 

Interestingly, both FI-based measures are shown to be insensitive to the 
change of symmetry, by varying the phase of squeezing $\varphi$ in Fig.\ref{fig:asymmetric}. Meanwhile, all the other measures have great dependence on the phase of squeezing $\varphi$. This is another convenient trait of FI-based measures of being tolerant to phase mismatch. As the phase of squeezing is usually determined by the specific experiment setups, such as the properties of cavity in cQED platform~\cite{leghtas2015confining} and nonlinear crystals in optical platform~\cite{dutt2015chip,schnabel2017squeezed}.

\section{Concluding remarks}\label{sec:conclusion}
In conclusion, this research provides a comprehensive analysis of quantum phase synchronization measures. Our work proposes a novel approach to measure the degree of synchronization by deploying classical and quantum Fisher information. Significantly, both measures demonstrate success in characterizing both the 1-to-1 and 2-to-1 synchronization regimes, where other existing methods fail to yield reliable results in one or another.

Our comparative study of the classical and quantum Fisher information measures with existing measures highlights the advantages and limitations of each method. Our study offers valuable guidance for future investigations and practical implementations. Our analysis of the impact of noise on the synchronization measures reveals the robustness and susceptibility of each method in the presence of  decoherence. Furthermore, the correlations between these measures provide insight into the similarities and differences between different measures of quantum synchronization.

Our findings contribute significantly to the characterization of quantum phase synchronization, particularly in the 2-to-1 synchronization regime. These results pave the way for further research in the field, such as the development of more efficient and robust quantum communication and computing protocols. Future work could explore other synchronization regimes, investigate the impact of various types of noise, and assess potential applications of our proposed measures in real-world quantum systems.

\section*{Acknowledgments}

Numerical simulations are performed using the QuTiP numerical toolbox~\cite{qutip1,qutip2}. YS and WJF would like to acknowledge the support from NRF-CRP19-2017-01, National Research Foundation, Singapore. LCK are grateful to the National Research Foundation, Singapore and the Ministry of Education, Singapore for financial support. 

\appendix

\section{Analytical solutions using density matrix ansatz}

In deep quantum regime ($\gamma_2\rightarrow \infty$), the $3\times3 $ density matrix ansatz in noiseless limit ($\gamma_3=0$) is given in  the Fock basis~\cite{mok2020sync_boost}:
\begin{equation}
    \rho = 
    \begin{pmatrix}
        \rho_{00} & \rho_{01} & 0 \\
        \rho_{10} &\rho_{11}& 0 \\
        0 &0 &\rho_{22}
    \end{pmatrix}
    ,
\end{equation}
with 
\begin{align}
    \rho_{00} &= \frac{\gamma_2(12E^2+18))}{12E^2+9+3\gamma_2(15+8E^2)},\\
    \rho_{11} &= \frac{\gamma_2(12E^2+9))}{12E^2+9+3\gamma_2(15+8E^2)},\\
    \rho_{22} &= \frac{12E^2+9}{12E^2+9+3\gamma_2(15+8E^2)},\\
    \rho_{01} &= \rho_{10}^{*} =\frac{6i \gamma_2 E}{12E^2+9+3\gamma_2(15+8E^2)}.
\end{align}

This amounts to restricting the number of excitations to $2$, and neglecting all coherences involving the state $\ket{2}$. The higher order coherences
are dropped on the grounds that they can be seen to be small in exact
numerical simulations, and dropping them makes analytical calculations
much easier and more insightful.

\begin{figure}[hbt]
    \centering
    \subfloat[]{
    \includegraphics[width=0.9\linewidth]{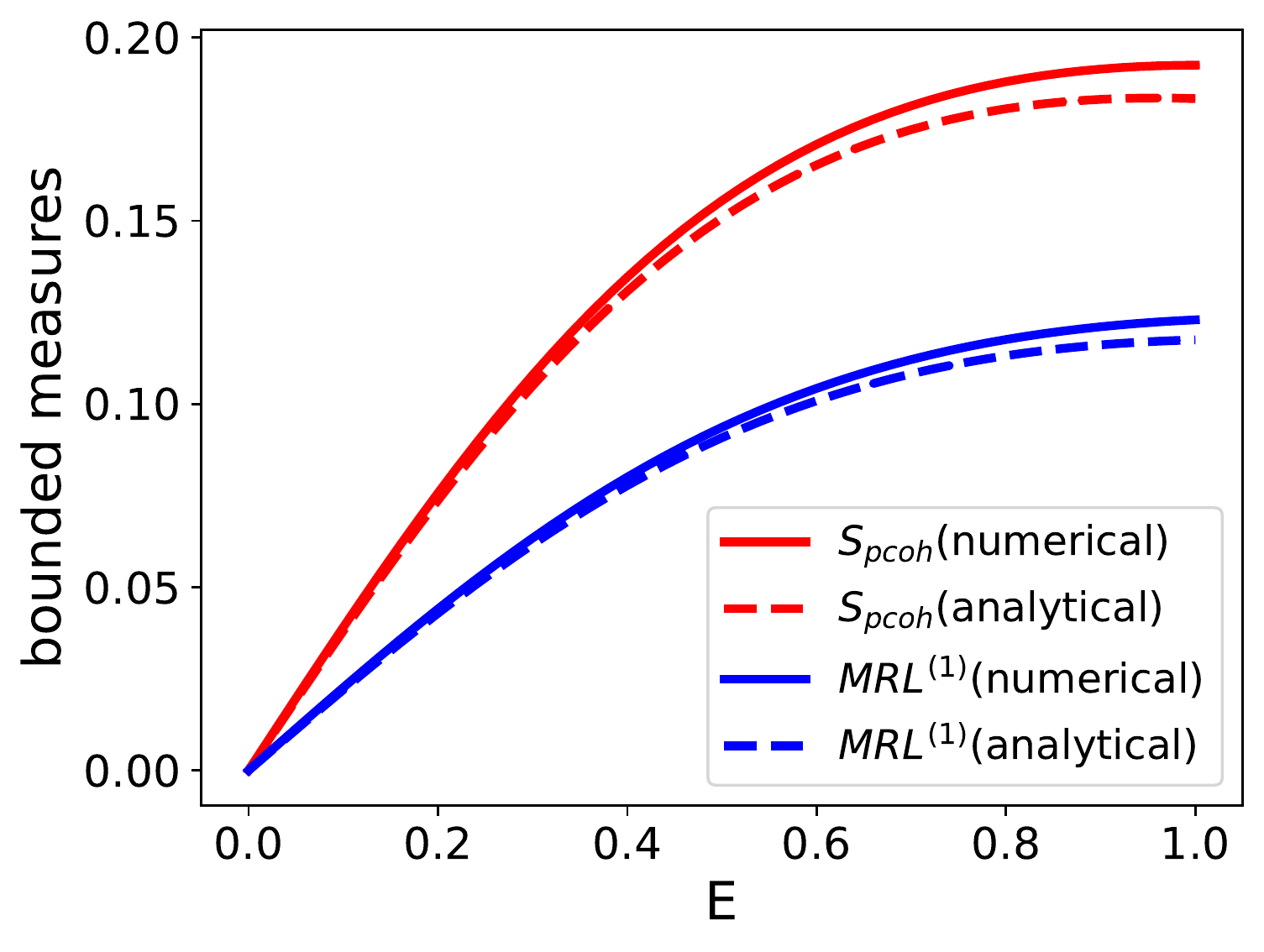}
    }

    \subfloat[]{
    \includegraphics[width=0.9\linewidth]{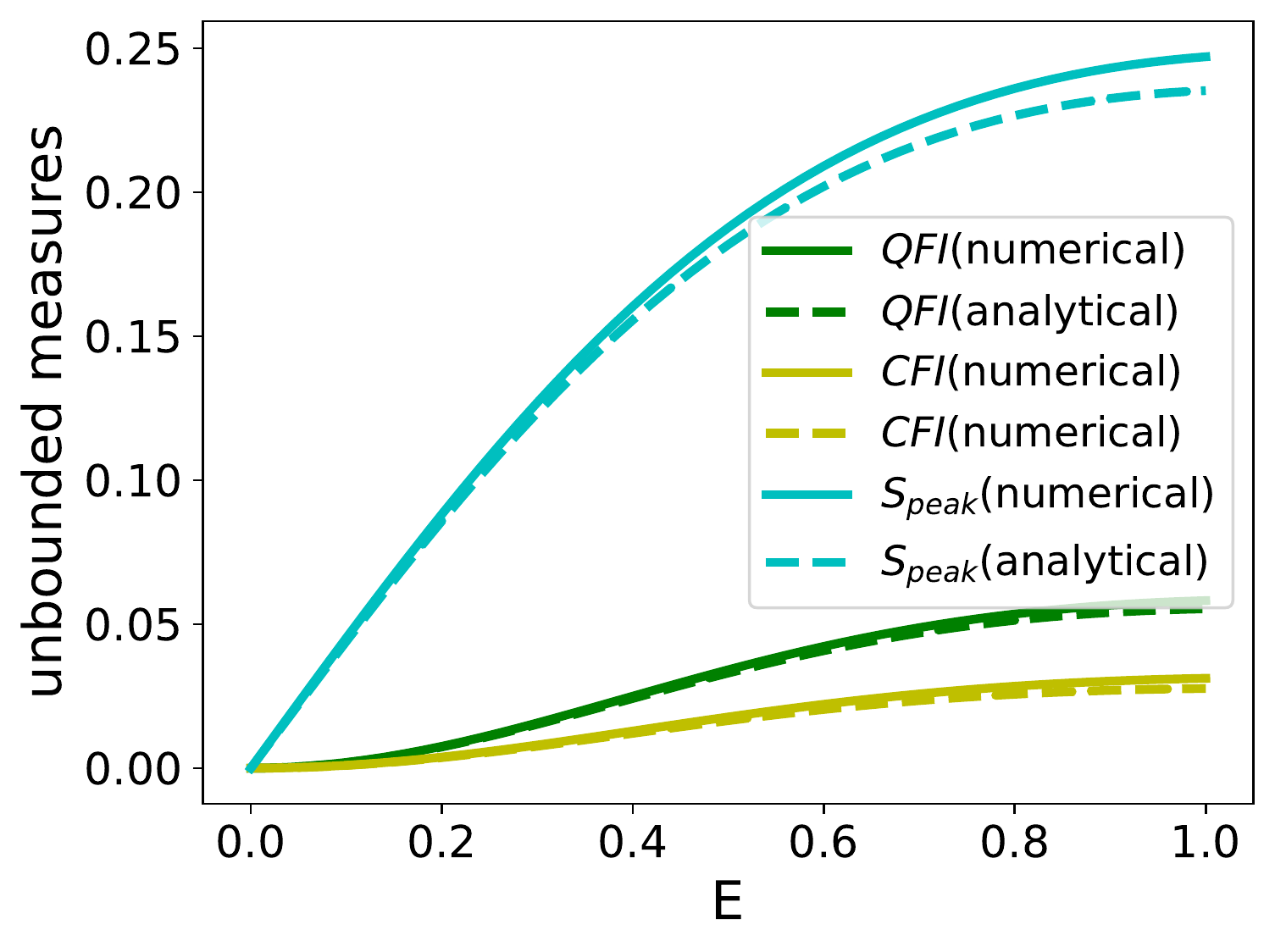}
    }
    \caption{Numerical results vs. analytical results. The oscillator is assumed in deep quantum regime ($\gamma_2=300$). Fixed parameters: $\Delta=0,\eta=0,\gamma_1=1,\gamma_3=0$.Analytical results are accurate only  when $E\ll 1$.}
    \label{fig:compare}
\end{figure}

In Fig.~\ref{fig:compare} we compare the analytical solutions to the numerical simulation, revealing that the validity of the solutions lies in the range when coherent driving is small ($E\ll 1$). With larger driving $E$, the oscillator will be excited to higher levels beyond the assumption of the $3\times3$ ansatz.

\bibliographystyle{unsrt} 
\bibliography{ref}

\begin{thebibliography}{10}

\bibitem{strogatz2012sync}
Steven~H Strogatz.
\newblock {\em Sync: How order emerges from chaos in the universe, nature, and
  daily life}.
\newblock Hachette UK, 2012.

\bibitem{pikovsky2001synchronization}
A~Pikovsky, M~Rosenblum, and J~Kurths.
\newblock {\em Synchronization Cambridge University Press}.
\newblock 2001.

\bibitem{strogatz2005crowd}
Steven~H Strogatz, Daniel~M Abrams, Allan McRobie, Bruno Eckhardt, and Edward
  Ott.
\newblock Crowd synchrony on the millennium bridge.
\newblock {\em Nature}, 438(7064):43--44, 2005.

\bibitem{prasad2010universal}
Awadhesh Prasad.
\newblock Universal occurrence of mixed-synchronization in counter-rotating
  nonlinear coupled oscillators.
\newblock {\em Chaos, Solitons \& Fractals}, 43(1-12):42--46, 2010.

\bibitem{czolczynski2012synchronization}
Krzysztof Czolczynski, Przemys{\l}aw Perlikowski, Andrzej Stefanski, and Tomasz
  Kapitaniak.
\newblock Synchronization of pendula rotating in different directions.
\newblock {\em Communications in Nonlinear Science and Numerical Simulation},
  17(9):3658--3672, 2012.

\bibitem{sharma2012counter-rotating}
Amit Sharma and Manish~Dev Shrimali.
\newblock {Experimental realization of mixed-synchronization in
  counter-rotating coupled oscillators}.
\newblock {\em Nonlinear Dynamics}, 69(1-2):371--377, 2012.
\newblock classical mixed sync.

\bibitem{sathiyadevi2022emerging}
K~Sathiyadevi, VK~Chandrasekar, and M~Lakshmanan.
\newblock Emerging chimera states under nonidentical counter-rotating
  oscillators.
\newblock {\em Physical Review E}, 105(3):034211, 2022.

\bibitem{zeng2011chaos}
Caibin Zeng, Qigui Yang, and Junwei Wang.
\newblock Chaos and mixed synchronization of a new fractional-order system with
  one saddle and two stable node-foci.
\newblock {\em Nonlinear Dynamics}, 65:457--466, 2011.

\bibitem{bhowmick2015counter-rotating}
Sourav~K. Bhowmick, Bidesh~K. Bera, and Dibakar Ghosh.
\newblock {Generalized counter-rotating oscillators: Mixed synchronization}.
\newblock {\em Communications in Nonlinear Science and Numerical Simulation},
  22(1-3):692--701, 2015.

\bibitem{bhowmick2012mixed}
Sourav~K Bhowmick, Chittaranjan Hens, Dibakar Ghosh, and Syamal~K Dana.
\newblock Mixed synchronization in chaotic oscillators using scalar coupling.
\newblock {\em Physics Letters A}, 376(36):2490--2495, 2012.

\bibitem{lorch2016phase-coherence}
Niels Lörch, Ehud Amitai, Andreas Nunnenkamp, and Christoph Bruder.
\newblock {Genuine Quantum Signatures in Synchronization of Anharmonic
  Self-Oscillators}.
\newblock {\em Physical Review Letters}, 117(7):073601, 2016.

\bibitem{lee2013quantum}
Tony~E Lee and HR~Sadeghpour.
\newblock Quantum synchronization of quantum van der {P}ol oscillators with
  trapped ions.
\newblock {\em Physical {R}eview {L}etters}, 111(23):234101, 2013.

\bibitem{walter2014quantum}
Stefan Walter, Andreas Nunnenkamp, and Christoph Bruder.
\newblock Quantum synchronization of a driven self-sustained oscillator.
\newblock {\em Physical Review Letters}, 112(9):094102, 2014.

\bibitem{walter2015quantum}
Stefan Walter, Andreas Nunnenkamp, and Christoph Bruder.
\newblock Quantum synchronization of two van der pol oscillators.
\newblock {\em Annalen der Physik}, 527(1-2):131--138, 2015.

\bibitem{jaseem2020generalized}
Noufal Jaseem, Michal Hajdu{\v{s}}ek, Parvinder Solanki, Leong-Chuan Kwek,
  Rosario Fazio, and Sai Vinjanampathy.
\newblock Generalized measure of quantum synchronization.
\newblock {\em Physical Review Research}, 2(4):043287, 2020.

\bibitem{mari2013measures}
Andrea Mari, Alessandro Farace, Nicolas Didier, Vittorio Giovannetti, and
  Rosario Fazio.
\newblock Measures of quantum synchronization in continuous variable systems.
\newblock {\em Physical Review Letters}, 111(10):103605, 2013.

\bibitem{sonar2018squeezing}
Sameer Sonar, Michal Hajdušek, Manas Mukherjee, Rosario Fazio, Vlatko Vedral,
  Sai Vinjanampathy, and Leong-Chuan Kwek.
\newblock {Squeezing Enhances Quantum Synchronization}.
\newblock {\em Physical Review Letters}, 120(16):163601, 2018.

\bibitem{weiss2016noise}
Talitha Weiss, Andreas Kronwald, and Florian Marquardt.
\newblock {Noise-induced transitions in optomechanical synchronization}.
\newblock {\em New Journal of Physics}, 18(1):013043, 2016.

\bibitem{frieden1990fisher}
B~Roy Frieden.
\newblock Fisher information, disorder, and the equilibrium distributions of
  physics.
\newblock {\em Physical Review A}, 41(8):4265, 1990.

\bibitem{frieden2004science}
B~Roy Frieden.
\newblock {\em Science from Fisher information: a unification}.
\newblock Cambridge University Press, 2004.

\bibitem{shen2023quantum_sync_nonlinear}
Yuan Shen, Wai-Keong Mok, Changsuk Noh, Ai~Qun Liu, Leong-Chuan Kwek, Weijun
  Fan, and Andy Chia.
\newblock Quantum synchronization effects induced by strong nonlinearities.
\newblock {\em Phys. Rev. A}, 107:053713, May 2023.

\bibitem{barak2005non}
R~Barak and Y~Ben-Aryeh.
\newblock Non-orthogonal positive operator valued measure phase distributions
  of one-and two-mode electromagnetic fields.
\newblock {\em Journal of Optics B: Quantum and Semiclassical Optics},
  7(5):123, 2005.

\bibitem{shen2023enhance_homodyne}
Yuan Shen, Hong~Yi Soh, Weijun Fan, and Leong-Chuan Kwek.
\newblock Enhancing quantum synchronization through homodyne measurement and
  squeezing, 2023.

\bibitem{hush2015spin_correlations}
Michael~R. Hush, Weibin Li, Sam Genway, Igor Lesanovsky, and Andrew~D. Armour.
\newblock Spin correlations as a probe of quantum synchronization in
  trapped-ion phonon lasers.
\newblock {\em Phys. Rev. A}, 91:061401, Jun 2015.

\bibitem{lorch2017sync_blockade}
Niels L\"orch, Simon~E. Nigg, Andreas Nunnenkamp, Rakesh~P. Tiwari, and
  Christoph Bruder.
\newblock Quantum synchronization blockade: Energy quantization hinders
  synchronization of identical oscillators.
\newblock {\em Phys. Rev. Lett.}, 118:243602, Jun 2017.

\bibitem{mok2020sync_boost}
W.-K. Mok, L.-C. Kwek, and H.~Heimonen.
\newblock Synchronization boost with single-photon dissipation in the deep
  quantum regime.
\newblock {\em Phys. Rev. Res.}, 2:033422, Sep 2020.

\bibitem{pewsey2013circular}
Arthur Pewsey, Markus Neuh{\"a}user, and Graeme~D Ruxton.
\newblock {\em Circular statistics in R}.
\newblock Oxford University Press, 2013.

\bibitem{kalloniatis2018fisher}
Alexander~C Kalloniatis, Mathew~L Zuparic, and Mikhail Prokopenko.
\newblock Fisher information and criticality in the kuramoto model of
  nonidentical oscillators.
\newblock {\em Physical Review E}, 98(2):022302, 2018.

\bibitem{da2021fisher}
VB~da~Silva, JP~Vieira, and Edson~D Leonel.
\newblock Fisher information of the kuramoto model: A geometric reading on
  synchronization.
\newblock {\em Physica D: Nonlinear Phenomena}, 423:132926, 2021.

\bibitem{yue2015operation}
Jie-Dong Yue, Yu-Ran Zhang, and Heng Fan.
\newblock Operation-triggered quantum clock synchronization.
\newblock {\em Physical Review A}, 92(3):032321, 2015.

\bibitem{zhang2013criterion}
Yong-Liang Zhang, Yu-Ran Zhang, Liang-Zhu Mu, and Heng Fan.
\newblock Criterion for remote clock synchronization with heisenberg-scaling
  accuracy.
\newblock {\em Physical Review A}, 88(5):052314, 2013.

\bibitem{jozsa2000quantum}
Richard Jozsa, Daniel~S Abrams, Jonathan~P Dowling, and Colin~P Williams.
\newblock Quantum clock synchronization based on shared prior entanglement.
\newblock {\em Physical Review Letters}, 85(9):2010, 2000.

\bibitem{chen2010clocks}
Ping Chen and Shunlong Luo.
\newblock Clocks and fisher information.
\newblock {\em Theoretical and Mathematical Physics}, 165:1552--1564, 2010.

\bibitem{steven1993fundamentals}
M~Kay Steven.
\newblock Fundamentals of statistical signal processing.
\newblock {\em PTR Prentice-Hall, Englewood Cliffs, NJ}, 10:151045, 1993.

\bibitem{paris2009quan_estimate}
MATTEO G.~A. PARIS.
\newblock {QUANTUM ESTIMATION FOR QUANTUM TECHNOLOGY}.
\newblock {\em International Journal of Quantum Information},
  7(supp01):125--137, 2009.

\bibitem{rath2021qfi_rand_meas}
Aniket Rath, Cyril Branciard, Anna Minguzzi, and Beno\^{\i}t Vermersch.
\newblock Quantum fisher information from randomized measurements.
\newblock {\em Phys. Rev. Lett.}, 127:260501, Dec 2021.

\bibitem{frowis2016detect_qfi}
Florian Fr\"owis, Pavel Sekatski, and Wolfgang D\"ur.
\newblock Detecting large quantum fisher information with finite measurement
  precision.
\newblock {\em Phys. Rev. Lett.}, 116:090801, Mar 2016.

\bibitem{yu2021exp_est_qfi}
Min Yu, Dongxiao Li, Jingcheng Wang, Yaoming Chu, Pengcheng Yang, Musang Gong,
  Nathan Goldman, and Jianming Cai.
\newblock Experimental estimation of the quantum fisher information from
  randomized measurements.
\newblock {\em Phys. Rev. Res.}, 3:043122, Nov 2021.

\bibitem{daniel2023geometric}
Aaron Daniel, Christoph Bruder, and Martin Koppenh{\"o}fer.
\newblock Geometric phase in quantum synchronization.
\newblock {\em arXiv preprint arXiv:2302.08866}, 2023.

\bibitem{tong2004kinematic}
DM~Tong, Erik Sj{\"o}qvist, Leong~Chuan Kwek, and Choo~Hiap Oh.
\newblock Kinematic approach to the mixed state geometric phase in nonunitary
  evolution.
\newblock {\em Physical review letters}, 93(8):080405, 2004.

\bibitem{ameri2015mutual}
V~Ameri, M~Eghbali-Arani, A~Mari, A~Farace, F~Kheirandish, V~Giovannetti, and
  R~Fazio.
\newblock Mutual information as an order parameter for quantum synchronization.
\newblock {\em Physical Review A}, 91(1):012301, 2015.

\bibitem{eneriz2019degree}
H~Eneriz, DZ~Rossatto, Francisco~A C{\'a}rdenas-L{\'o}pez, E~Solano, and
  M~Sanz.
\newblock Degree of quantumness in quantum synchronization.
\newblock {\em Scientific Reports}, 9(1):19933, 2019.

\bibitem{leghtas2015confining}
Z.~Leghtas, S.~Touzard, I.~M. Pop, A.~Kou, B.~Vlastakis, A.~Petrenko, K.~M.
  Sliwa, A.~Narla, S.~Shankar, M.~J. Hatridge, M.~Reagor, L.~Frunzio, R.~J.
  Schoelkopf, M.~Mirrahimi, and M.~H. Devoret.
\newblock Confining the state of light to a quantum manifold by engineered
  two-photon loss.
\newblock {\em Science}, 347(6224):853--857, 2015.

\bibitem{dutt2015chip}
Avik Dutt, Kevin Luke, Sasikanth Manipatruni, Alexander~L Gaeta, Paulo
  Nussenzveig, and Michal Lipson.
\newblock On-chip optical squeezing.
\newblock {\em Physical Review Applied}, 3(4):044005, 2015.

\bibitem{schnabel2017squeezed}
Roman Schnabel.
\newblock Squeezed states of light and their applications in laser
  interferometers.
\newblock {\em Physics Reports}, 684:1--51, 2017.

\bibitem{qutip1}
J.R. Johansson, P.D. Nation, and Franco Nori.
\newblock Qutip: An open-source python framework for the dynamics of open
  quantum systems.
\newblock {\em Computer Physics Communications}, 183(8):1760--1772, 2012.

\bibitem{qutip2}
J.R. Johansson, P.D. Nation, and Franco Nori.
\newblock Qutip 2: A python framework for the dynamics of open quantum systems.
\newblock {\em Computer Physics Communications}, 184(4):1234--1240, 2013.

\end{thebibliography}

\end{document}